\documentclass[twocolumn,aps,prl]{revtex4}

\begin{document}

%\title{}

%\maketitle

%\newpage
\narrowtext

{\bf Comment on ``Strongly correlated Fractional Quantum Hall line
junction''}

\vspace{1ex}

Letter \cite{b1} and subsequent extended paper proposed an ``exact
solution'' of the problem of a tunnel junction of length $L$ between
two single-mode edges of the Fractional Quantum Hall Liquids with
different filling factors $\nu_{j}=1/(2m_j+1)$, $j=1,2$;
$m_j=0,1,...$. The authors use a special choice of parameters which
makes possible \cite{b3,b4} simple description of the tunneling part
of such a ``line'' junction using fermionization, and argue that
this choice is justified in the limit of infinitely strong but local
Coulomb interaction of the closely-spaced edges. The purpose of this
comment is to point out that the solution found in \cite{b1} is not
correct and to present correct solution of a particular instance of
this model. One important consequence of our solution is that the
maximum tunneling conductance $G$ of the line junction has the same
``strong-coupling'' value $G=G_m\equiv 2\nu_1 \nu_2/(\nu_1+\nu_2)$
(in units of the free electron conductance $\sigma=e^2/h$) as a
point contact \cite{b5}, and not the smaller value $G_J= \mbox{min}
\{\nu_{m_1},\nu_{m_2}\}$ obtained in \cite{b1}.

The mistake in \cite{b1} is caused by unjustified assumption of the
existence of the local chemical potentials not only for the incoming
but also for the outgoing edges. This assumption is important since
the junction conductance $G$ is calculated in \cite{b1} by matching
these potentials across the ends of the junction. Existence of the
chemical potentials implies that the local equilibrium is imposed at
the junction end points $x=\pm L/2$, while actually there is no
equilibrium: the chemical potentials are defined only for the
incoming edges, and all the rest follows from coherent quantum
evolution of the fields governed by the edge Hamiltonian \cite{b6}.

Correct matching between the junction and external edges
consists in imposing the continuity of the edge Bose
fields $\phi_j(x,t)$ at $x=\pm L/2$ and coincides with
the standard "unfolded" form \cite{b7} of
multi-component Dirichlet boundary condition. The fields
$\phi_{1,2}$ are normalized so that the edge currents
are $j_i=-e \sqrt{\nu_i}\partial_t \phi_i/2\pi$. Under
the transformation \cite{b3,b4,b1} of $\phi_{1,2}$ into
the ``charge'' and ``spin'' modes $\phi_{c,n} (x,t)$
propagating independently inside the junction $|x|<L/2$,
the continuity conditions  in the notations of \cite{b1}
have the following matrix form:\\
\hspace*{2cm} $ \phi_{c,n}(x,t) = M
\phi_{1,2}(x,t)|_{x=\pm L/2}\, ,$
\begin{equation}
M= \frac{1}{\sqrt{\nu_1-\nu_2}} \left(
\begin{array}{cc} \displaystyle \sqrt{\nu_1} \, , & \sqrt{\nu_2}
\\ \sqrt{\nu_2} \, , & \sqrt{\nu_1} \end{array} \right)
\, . \label{e1}
\end{equation}
taking $\nu_1 > \nu_2$. The charge mode $\phi_c (x-v_ct)$ is a free
chiral field which moves with some velocity $v_c$ in the same,
``positive'', direction as $\phi_1$. Hence, the field values at the
end points are related as $\phi_c (L/2,t+L/v_c)=\phi_c (-L/2,t)$.
The spin mode $\phi_n(x)$ has the opposite chirality. Its dynamics
is affected by tunneling and can be solved by refermionization
\cite{b3,b4}. We assume that the tunnel amplitude $\Delta$ is
constant throughout the junction, and limit our discussion here to
the simplest case $m_2-m_1=1$, when the fermionic representation of
the junction dynamics \cite{b3} has the form of the
tunneling-induced rotation between two components of a Dirac fermion
propagating with a velocity $v_n$. The spin mode is defined by the
difference between the density operators of the two components.

Although this fermionic representation does not provide a general
simple relation between the spin mode operators at the boundaries,
for special values of the junction length: $L=(\pi v_n/2\Delta )l$,
where $l=1,2...$, it shows that $\phi_n (-L/2,t+L/v_n)=(-1)^l \phi_n
(L/2,t)$. Combined with Eq.~(\ref{e1}), this relation gives complete
description of the dynamics of the fields $\phi$ in the line
junction. This dynamics exhibits multiple interference governed by
the reflection amplitudes $\sqrt{\nu_2/\nu_1}$ at the end points
resembling \cite{b8} a Tomonaga-Luttinger wire connected to leads.

To find the zero-frequency conductance $G$, one can ignore the
finite times of propagation between the end points. This means that
for even $l$ the fields $\phi_{1,2}$ are not changed and $G=0$. For
odd $l$, the spin mode changes sign and the total transformation $T$
of the fields $\phi_{1,2}$ is:
\begin{equation}
\phi_{1,2}|_{x=L/2} = T \phi_{1,2}|_{x=-L/2}\, , \;\;\; T=M^{-1}
\sigma_z M \, . \label{e2}
\end{equation}
Using Eq.~(\ref{e1}), and changing the transfer matrix $T$
(\ref{e2}) into the scattering matrix $P$ which relates incoming and
outgoing fields $\phi$, one can immediately see that $P$ coincides
with the scattering matrix of a point contact \cite{b5} (or in fact
any odd number of successive point contacts \cite{b9}) in the
strong-coupling limit. In this case the junction conductance is
$G=G_m$ contrary to the result obtained in \cite{b1} (see Eqs.~(14)
and (40)-(42) of, respectively, the short and long papers).
Qualitatively, based on the nature of the fermionic dynamics inside
the junction, one can expect $G$ to oscillate with $L$ between the
maximum $G_m$ and the minimum $G=0$. \vspace{0.4cm}

\noindent
Vadim V. Ponomarenko\\
\indent International Center for Condensed Matter Physics,\\
\indent Universidade de Brasília, 70910-900 Brasília, Brazil,\\
\indent and Center for Advanced Studies, \\
\indent St. Petersburg State Polytechnical University,\\
\indent  St. Petersburg 195251, Russia.

\noindent
Dmitri V. Averin \\
\indent Department of Physics and Astronomy \\
\indent Stony Brook University, SUNY\\
\indent Stony Brook, NY 11794-3800, USA

\end{document}